# Generalized electromechanical stability of dielectric elastomers

Fengxian Xin[1,2,*] and Tian Jian Lu[1,2,∗]

[1]*MOE Key Laboratory for Multifunctional Materials and Structures,*

*Xi'an Jiaotong University, Xi'an 710049, P.R. China*

[2]*State Key Laboratory for Strength and Vibration of Mechanical Structures,*

*Xi'an Jiaotong University, Xi'an 710049, P.R. China*

We investigate the generalized electromechanical stability of dielectric elastomer actuators by considering the competition between Maxwell and mechanical stresses. Existing researches based on the positive definiteness of the nominal Hessian matrix addressed the initial load routine stability, i.e., Born stability, in the absence of prestresses. We generalize this method to account for electromechanical stability at any prestress state, which fully takes into account the pull-in instability and the thermodynamic instability (in the case of electromechanical system). The given stability phase diagrams clearly define the boundaries of different instability regions. The generalized method for electromechanical stability can help the factual design of actuators constructed from dielectric elastomers.

## I. INTRODUCTION

A layer of dielectric elastomer subjected to electric voltage can reduce its thickness and expand its area, ideal for actuators[1-4]. However, a voltage-controlled actuator often possesses a nonconvex free energy function (combination of elastic strain energy and electric field energy), thus susceptible to electromechanical instability that manifests as either pull-in instability or inhomogeneous large deformation[5]. Although the electromechanical instability of dielectric elastomers has been extensively studied using the positive definiteness of the nominal Hessian matrix[5-7], this approach is mainly applicable for Born stability, namely, initial load route stability[8-10]. Judgment of electromechanical stability should be evaluated in the true space (deformed state), not the nominal space (undeformed state): the positive definiteness of the nominal Hessian matrix holds in the special case of zero prestress, because only under such

---

conditions (*i.e.*, no prestress and no deformation) the nominal space is the true space[8-13].

As an extension of our recent work[14], here we aim to present the generalized stability criteria for a layer of dielectric elastomer sandwiched between two coated electrodes, which takes account of both the pull-in instability and thermodynamic instability. Specifically, we provide critical border lines between pull-in stability-instability and electromechanical stability-instability for this type of electromechanical system. We further demonstrate that the existing method for analyzing the electromechanical stability of dielectric elastomers can give approximate predictions in the presence of prestresses.

## II. THEORETICAL MODEL

Consider a thin layer of dielectric elastomer coated with two compliant electrodes, with dimensions ($L_1$, $L_2$, $L_3$) in undeformed state. Upon applying a voltage $\Phi$ on the electrodes, the sandwich deforms to current dimensions ($l_1$, $l_2$, $l_3$), with principal stretches ($\lambda_1$, $\lambda_2$, $\lambda_3$). Since the elastomer can undergo large shape changes with little volume change, it is assumed incompressible, requiring $\lambda_1 \lambda_2 \lambda_3 = 1$. The electric field in current state is $E = \Phi \lambda_1 \lambda_2 / L_3$ and the charge on either electrode is $Q = L_1 L_2 \varepsilon \Phi \lambda_1^2 \lambda_2^2 / L_3$, $\varepsilon$ being permittivity. The elastomer layer is subjected to Maxwell stresses ($-\varepsilon E^2/2$, $-\varepsilon E^2/2$, $\varepsilon E^2/2$) which, due to incompressibility, are equivalent to equal-biaxial stresses ($-\varepsilon E^2$, $-\varepsilon E^2$, 0) or uniaxial stress (0, 0, $\varepsilon E^2$). The two stress states of equal-biaxial prestresses and uniaxial prestress will be analyzed below.

The nonlinear elastic behavior of dielectric elastomer with small-strain shear modulus $\mu$ can be modeled by adopting the free energy function[15], as:

$$W_s(\lambda_1, \lambda_2) = \frac{\mu}{2}\left(\lambda_1^2 + \lambda_2^2 + \lambda_1^{-2}\lambda_2^{-2} - 3\right) \tag{1}$$

It follows that $\sigma_i = \lambda_i \, \partial W_s / \partial \lambda_i$ is the true stress (Cauchy stress) induced by elastic deformation. The electromechanical response of the dielectric elastomer is intrinsically governed by three independent variables: voltage $\Phi$, in-plane stretches $\lambda_1$ and $\lambda_2$. As the electromechanical stability of the system is voltage controlled, its constitutive relation is expressed in terms of the three basic variables as:

$$\sigma_1 = \mu\left(\lambda_1^2 - \lambda_1^{-2}\lambda_2^{-2}\right) - \varepsilon\left(\frac{\Phi}{L_3}\right)^2 \lambda_1^2 \lambda_2^2 \tag{2}$$

$$\sigma_2 = \mu\left(\lambda_2^2 - \lambda_1^{-2}\lambda_2^{-2}\right) - \varepsilon\left(\frac{\Phi}{L_3}\right)^2 \lambda_1^2 \lambda_2^2 \tag{3}$$

$$Q = \frac{L_1 L_2 \varepsilon \Phi}{L_3} \lambda_1^2 \lambda_2^2 \tag{4}$$

The stability of a finite deformed elastic body is governed by strong ellipticity condition, as:

$$(\mathbf{b} \otimes \mathbf{N}) : \frac{\partial^2 W}{\partial \mathbf{F} \partial \mathbf{F}} : (\mathbf{b} \otimes \mathbf{N}) > 0 \tag{5}$$

for arbitrary $\mathbf{b} \otimes \mathbf{N} \neq 0$, where $\mathbf{F}$ is the deformation gradient. The nominal Hessian matrix is defined accordingly as $\mathbf{H} \equiv \frac{\partial^2 W}{\partial \mathbf{F} \partial \mathbf{F}}$. The voltage or the Maxwell stress can be considered as part of the material law, thus the free energy should be the strain energy minus the work done by the outside voltage source, as $W(\lambda_1, \lambda_2, \Phi) = \frac{\mu}{2}\left(\lambda_1^2 + \lambda_2^2 + \lambda_1^{-2}\lambda_2^{-2} - 3\right) - \frac{1}{2}\varepsilon\left(\frac{\Phi}{L_3}\right)^2 \lambda_1^2 \lambda_2^2$. In Eulerian coordinates, condition (6) can be rewritten as[8-10]:

$$(\mathbf{b} \otimes \mathbf{n}) : (\mathbf{C} + \boldsymbol{\sigma} \otimes \mathbf{I}) : (\mathbf{b} \otimes \mathbf{n}) > 0 \tag{7}$$

where $\mathbf{n} = \mathbf{F}^{-T}\mathbf{N}$, $\mathbf{I}$ is the identity matrix, $\boldsymbol{\sigma}$ is the Cauchy stress, and $\mathbf{C} = \frac{1}{\det(\mathbf{F})} \mathbf{F} \frac{\partial^2 W}{\partial \mathbf{F} \partial \mathbf{F}} \mathbf{F}^{T}$. Therefore, stability of any elastic system is determined by the positive definiteness of the true tangential stiffness matrix $\mathbf{c} \equiv (\mathbf{C} + \boldsymbol{\sigma} \otimes \mathbf{I})$. In the special case of zero prestress, this condition can be degraded to the positive definiteness of initial tangential stiffness matrix $\mathbf{C}$ or the nominal Hessian matrix $\mathbf{H}$ that has been extensively applied to analyze the electromechanical stability of dielectric elastomers[5-7,16-18]. However, the positive definiteness of $\mathbf{H}$ can only be used to judge the stability of initial load route point (i.e., the Born stability) and is invalid if any prestress is present [8-10].

For any given $\Phi$, we can picture a total energy surface determined by ($\lambda_1$, $\lambda_2$), so that the true stress function $\sigma_2(\sigma_1)$ or nominal stress function $s_2(s_1)$ defines the load route or direction. At the initial load route point ($\sigma_1 = \sigma_2 = 0$), there exists no prestress and the convexity of the energy surface is determined by $\mathbf{H}$ (=$\mathbf{C}$). For arbitrary prescribed load route $\sigma_2(\sigma_1)$ or $s_2(s_1)$, the corresponding nominal Hessian matrix $\mathbf{H}$ describes the convexity of energy surface along the direction of this load route at the zero prestress point. From this viewpoint, previous investigations based on **H** described only the stability at initial load route point along direction $\sigma_2(\sigma_1)$ or $s_2(s_1)$. The generalized stability at arbitrary prestress state is determined by the positive definiteness of the true tangential stiffness matrix **c**, which is degraded to the positive definiteness of **H** (=**C**) when $\boldsymbol{\sigma}=0$.

The generalized constitutive relation of the considered electromechanical system may be expressed as $[d\sigma_1 \quad d\sigma_2 \quad dQ]^\mathrm{T} = [\mathbf{c}][d\varepsilon_1 \quad d\varepsilon_2 \quad d\Phi]^\mathrm{T}$. The true tangential stiffness matrix is given by:

$$\mathbf{c} = \begin{bmatrix} \lambda_1 \dfrac{\partial \sigma_1}{\partial \lambda_1} & \lambda_2 \dfrac{\partial \sigma_1}{\partial \lambda_2} & \dfrac{\partial \sigma_1}{\partial \Phi} \\ \lambda_1 \dfrac{\partial \sigma_2}{\partial \lambda_1} & \lambda_2 \dfrac{\partial \sigma_2}{\partial \lambda_2} & \dfrac{\partial \sigma_2}{\partial \Phi} \\ \lambda_1 \dfrac{\partial Q}{\partial \lambda_1} & \lambda_2 \dfrac{\partial Q}{\partial \lambda_2} & \dfrac{\partial Q}{\partial \Phi} \end{bmatrix} \tag{8}$$

which, in view of Eqs. (2)-(4), becomes:

$$\mathbf{c} = \begin{bmatrix} 2\mu(\lambda_1^2 + \lambda_1^{-2}\lambda_2^{-2}) - 2\varepsilon\left(\dfrac{\Phi}{L_3}\right)^2 \lambda_1^2\lambda_2^2 & 2\mu\lambda_1^{-2}\lambda_2^{-2} - 2\varepsilon\left(\dfrac{\Phi}{L_3}\right)^2 \lambda_1^2\lambda_2^2 & -2\varepsilon\left(\dfrac{\Phi}{L_3^2}\right)\lambda_1^2\lambda_2^2 \\ 2\mu\lambda_1^{-2}\lambda_2^{-2} - 2\varepsilon\left(\dfrac{\Phi}{L_3}\right)^2 \lambda_1^2\lambda_2^2 & 2\mu(\lambda_2^2 + \lambda_1^{-2}\lambda_2^{-2}) - 2\varepsilon\left(\dfrac{\Phi}{L_3}\right)^2 \lambda_1^2\lambda_2^2 & -2\varepsilon\left(\dfrac{\Phi}{L_3^2}\right)\lambda_1^2\lambda_2^2 \\ 2\dfrac{L_1 L_2 \varepsilon \Phi}{L_3}\lambda_1^2\lambda_2^2 & 2\dfrac{L_1 L_2 \varepsilon \Phi}{L_3}\lambda_1^2\lambda_2^2 & \dfrac{L_1 L_2 \varepsilon}{L_3}\lambda_1^2\lambda_2^2 \end{bmatrix} \tag{9}$$

This matrix should be positive definite at equilibrium state.

## III. RESULTS AND DISCUSSION

*3.1. Equal-biaxial prestresses*

Consider first the particular case of equal-biaxial prestresses, with $\sigma_1 = \sigma_2$ and $\lambda_1 = \lambda_2$. The pull-in instability appears when $\sigma_1 = \sigma_2 = 0$, and the stretch limit satisfies:

$$\frac{\Phi}{L_3\sqrt{\mu/\varepsilon}} = \sqrt{\lambda^{-2} - \lambda^{-8}} \tag{10}$$

The inequality

$$\frac{\Phi}{L_3\sqrt{\mu/\varepsilon}} < \sqrt{\lambda^{-2} - \lambda^{-8}} \tag{11}$$

determines the region in the non-dimensional plane of ($\lambda$, $\Phi/(L_3\sqrt{\mu/\varepsilon})$) where stability is warranted. To compare with existing results[7], inequality (12) can be rewritten in another non-dimensional space ($\tilde{D}/\sqrt{\mu\varepsilon}$, $\lambda$), as:

$$\lambda^6 > \left(\frac{\tilde{D}}{\sqrt{\mu\varepsilon}}\right)^2 + 1 \tag{13}$$

where $\tilde{D} = Q/(L_1 L_2)$ is the nominal electrical displacement.

A stable electromechanical system requires its true tangential stiffness matrix to be positive definite, as:

$$\begin{vmatrix} \lambda\dfrac{\partial \sigma}{\partial \lambda} & \dfrac{\partial \sigma}{\partial \Phi} \\ \lambda\dfrac{\partial Q}{\partial \lambda} & \dfrac{\partial Q}{\partial \Phi} \end{vmatrix} > 0 \tag{14}$$

or, more specifically, as:

$$\begin{vmatrix} \mu(2\lambda^2 + 4\lambda^{-4}) - 4\varepsilon\left(\dfrac{\Phi}{L_3}\right)^2 \lambda^4 & -2\varepsilon\left(\dfrac{\Phi}{L_3^2}\right)\lambda^4 \\ 4\dfrac{L_1 L_2 \varepsilon \Phi}{L_3}\lambda^4 & \dfrac{L_1 L_2 \varepsilon}{L_3}\lambda^4 \end{vmatrix} > 0 \tag{15}$$

All the principal minors of the above determinant should be positive, namely,

$$\mu(2\lambda^2 + 4\lambda^{-4}) - 4\varepsilon\left(\frac{\Phi}{L_3}\right)^2 \lambda^4 > 0, \quad \frac{L_1 L_2 \varepsilon}{L_3}\lambda^4 > 0 \text{ and } \Delta = \frac{L_1 L_2 \varepsilon \mu}{L_3}\left[2\lambda^6 + 4 + 4\left(\frac{\Phi}{L_3\sqrt{\mu/\varepsilon}}\right)^2 \lambda^8\right] > 0.$$

Finally, the electromechanical stability condition is given by:

$$\frac{\Phi}{L_3\sqrt{\mu/\varepsilon}} < \sqrt{\frac{\lambda^{-2} + 2\lambda^{-8}}{2}} \tag{16}$$

which is equivalent to the following inequality expressed in terms of nominal electrical displacement, as:

$$\lambda^6 > 2\left(\frac{\tilde{D}}{\sqrt{\mu\varepsilon}}\right)^2 - 2 \tag{17}$$

For the electromechanical system considered, eqs. (11) and (16) present the condition of pull-in stability while eqs. (13) and (17) ensure electromechanical stability. These results for equal-biaxial prestresses are different from those of existing studies [7], since the latter only described the stability at zero stress point along the direction of equal-biaxial load route.

Figure 1 plots the predicted stability phase diagram of dielectric elastomers subjected to equal-biaxial prestress in parameter space ($\lambda_1$, $\Phi/(L_3\sqrt{\mu/\varepsilon})$). Region A below the two border lines is the completely stable domain. In region B, pull-instability occurs but electromechanical instability is avoided. In region C, electromechanical instability occurs but pull-in instability does not. In region D, the voltage is large enough to induce both pull-in instability and electromechanical instability. To facilitate comparison with existing studies [7], our predictions are also plotted in parameter space ($\tilde{D}/\sqrt{\mu\varepsilon}$, $\lambda_1$), as shown in Fig. 2. Existing studies based on the nominal Hessian matrix can give approximate predictions of electromechanical stability in small stretch state, which however gradually diverge from the generalized results in large stretch state.

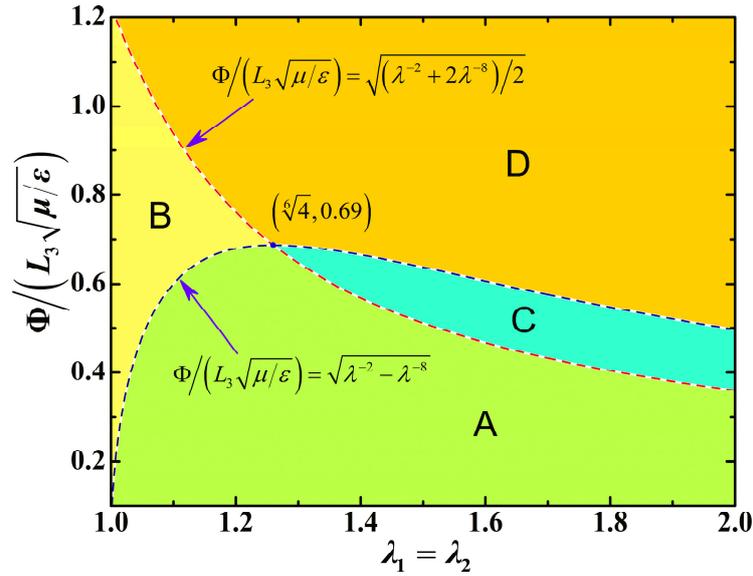

FIG. 1 (Color online) Stability phase diagram of dielectric elastomers subjected to equal-biaxial prestresses: A. completely stable region; B. electromechanically stable but pull-in instable region; C. pull-in stable but electromechanically instable region; D. electromechanical instable and pull-in instable region.

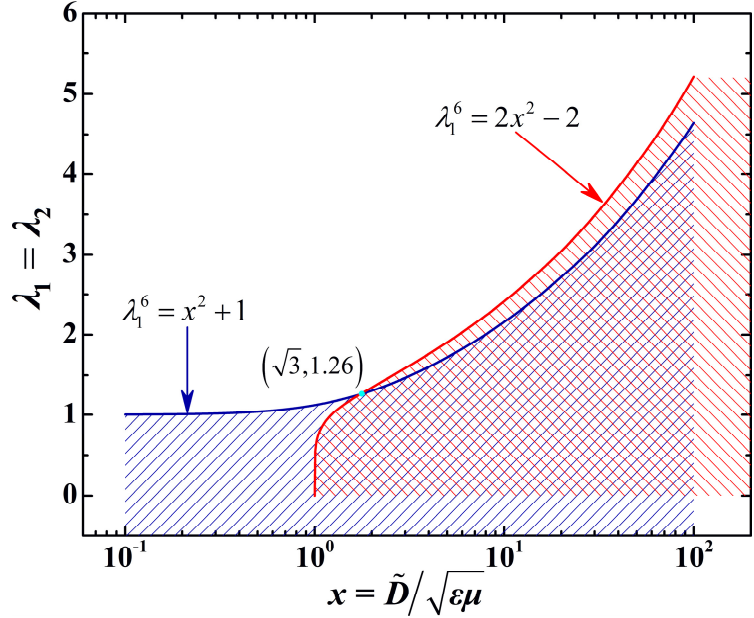

FIG. 2 (Color online) Stability phase diagram of dielectric elastomers subjected to equal-biaxial prestresses. Blue line is border line between pull-in stability and instability and red line is border line between electromechanical stability and instability. Shaded regions represent corresponding instable areas.

*3.2 Uniaxial prestresses*

Adopting similar procedures, we analyze next the stability of dielectric elastomers subjected to uniaxial prestress in the thickness direction, so that $\sigma_3 = \sigma$, $\lambda_3 = \lambda$ and $\lambda_1 = \lambda_2 = \lambda^{-1/2}$. The constitutive equations simplify to:

$$\sigma = \mu\left(\lambda^2 - \lambda^{-1}\right) + \varepsilon\left(\frac{\Phi}{L_3}\right)^2 \lambda^{-2} \tag{18}$$

$$Q = \frac{L_1 L_2 \varepsilon \Phi}{L_3} \lambda^{-2} \tag{19}$$

The critical condition for the occurrence of pull-in instability is $\sigma = 0$, and any further increase of voltage will cause the pull-in instability. Therefore, to ensure pull-in stability, the voltage should satisfy:

$$\frac{\Phi}{L_3\sqrt{\mu/\varepsilon}} < \sqrt{\lambda - \lambda^4} \tag{20}$$

or, equivalently, the stretch should satisfy:

$$\lambda^3 < \frac{1}{\left(\tilde{D}/\sqrt{\mu\varepsilon}\right)^2 + 1} \tag{21}$$

On the other hand, electromechanical stability dictates the positive definiteness of the matrix:

$$\begin{vmatrix} \mu\left(2\lambda^2 + \lambda^{-1}\right) - 2\varepsilon\left(\frac{\Phi}{L_3}\right)^2 \lambda^{-2} & 2\varepsilon\left(\frac{\Phi}{L_3^2}\right)\lambda^{-2} \\ -2\frac{L_1 L_2 \varepsilon \Phi}{L_3}\lambda^{-2} & \frac{L_1 L_2 \varepsilon}{L_3}\lambda^{-2} \end{vmatrix} > 0 \tag{22}$$

from which we need to ensure the following inequalities: $\mu\left(2\lambda^2 + \lambda^{-1}\right) - 2\varepsilon\left(\frac{\Phi}{L_3}\right)^2 \lambda^{-2} > 0$,

$\frac{L_1 L_2 \varepsilon}{L_3}\lambda^{-2} > 0$ and $\Delta = \frac{L_1 L_2 \varepsilon \mu}{L_3}\left[2 + \lambda^{-3} + 2\left(\frac{\Phi}{L_3\sqrt{\mu/\varepsilon}}\right)^2 \lambda^{-4}\right] > 0$. It follows that:

$$\frac{\Phi}{L_3\sqrt{\mu/\varepsilon}} < \sqrt{\frac{2\lambda^4 + \lambda}{2}} \tag{23}$$

which is equivalent to:

$$\lambda^3 < \frac{1}{2\left(\tilde{D}/\sqrt{\mu\varepsilon}\right)^2 - 2} \tag{24}$$

Figure 3 presents the stability phase diagram for the case of uniaxial prestress. Region A is the completely stable phase, region B is the pull-in instable phase, and region C is the electromechanical instable phase. In region D, both the pull-in instability and electromechanical instability are controlled. Again, these results are re-plotted in Fig. 4 in order to compare with existing studies [7]. The relatively large voltage in the shaded area will cause the corresponding instability. Predictions obtained previously using the positive definiteness of the nominal Hessian matrix approximately match with the present generalized results when the stretch (thinning in thickness direction) is small.

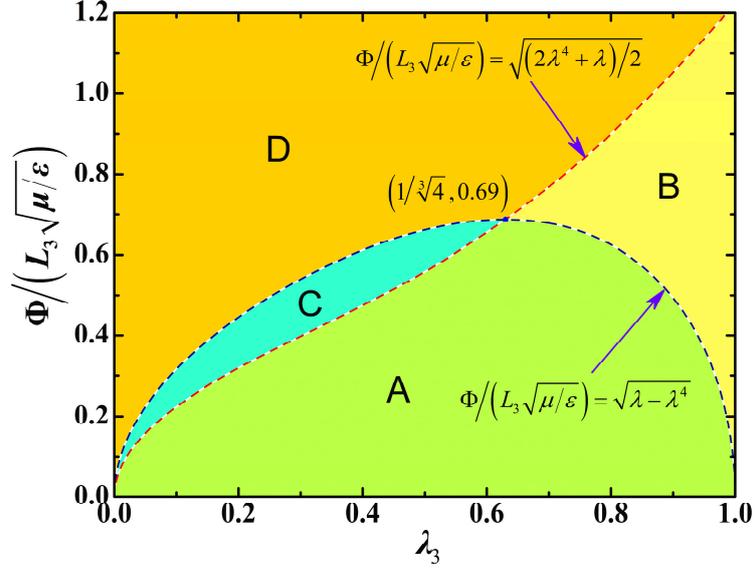

FIG. 3 (Color online) Stability phase diagram of dielectric elastomers under uniaxial prestress in thickness direction: A. completely stable region; B. electromechanical stable but pull-in instable region; C. pull-in stable but electromechanical instable region; D. electromechanical instable and pull-in instable region.

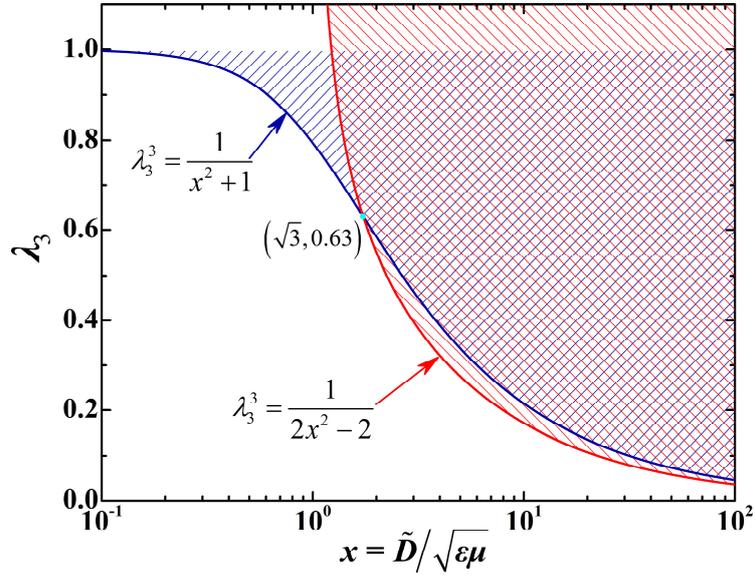

FIG. 4 (Color online) Stability phase diagram of dielectric elastomers under uniaxial prestress in thickness direction. Blue line is border line between pull-in stability and instability and red line is border line between electromechanical stability and instability. Shaded regions represent corresponding instable regimes.

The critical points appearing in the two phase diagrams of Figs. 1 and 3 correspond to the same electrical voltage of $\Phi/\left(L_3\sqrt{\mu/\varepsilon}\right) \approx 0.69$. This is understandable since, for incompressible materials, the state of equal-biaxial stresses in plane is intrinsically the same as that of uniaxial stress in the thickness direction. Therefore, the critical voltage should be identical for the two prestress cases considered.

## IV. CONCLUSIONS

We propose generalized criteria for avoiding pull-in instability and electromechanical instability in dielectric elastomer based electromechanical systems. The latter is based on the convexity of total system energy, which can be ensured by enforcing the positive definiteness of the true tangential stiffness matrix. The commonly applied nominal Hessian matrix can be employed to judge the stability in zero-prestress cases. In the presence of relatively large prestresses, predictions based on the nominal Hessian matrix may deviate significantly from the generalized results. The present results clearly define the boundaries between the stable region, pull-in instable region and the electromechanical instable region, which are expected to be helpful for the design of electromechanical actuators made of dielectric elastomers.

## ACKNOWLEDGEMENTS

This work was supported by the National Natural Science Foundation of China (51528501 and 11321062) and the Fundamental Research Funds for Central Universities (2014qngz12). F.X. Xin is supported by China Scholarship Council as a visiting scholar to Harvard University. This author thanks the helpful discussion with Prof. Zhigang Suo at Harvard University for the soft material theory.